\documentclass[prc, twocolumn]{revtex4}
\usepackage{amsmath}
\usepackage{graphicx}
\usepackage{xcolor}
\def\ra{\rangle}
\def\la{\langle}

\def\bx{{\bf{x}}}
\def\bq{{\bf{q}}}
\newcommand{\bJ}{{\bf{J}}}
\newcommand{\bD}{{\bf{D}}}
\newcommand{\bY}{{\bf{Y}}}
\def\bnabla{{\bf{\nabla}}}

\def\ra{\rangle}
\def\la{\langle}
\newskip\humongous \humongous=0pt plus 1000pt minus 1000pt
\def\caja{\mathsurround=0pt}

\newif\ifdtup
\def\panorama{\global\dtuptrue \openup1\jot \caja
        \everycr{\noalign{\ifdtup \global\dtupfalse
        \vskip-\lineskiplimit \vskip\normallineskiplimit
        \else \penalty\interdisplaylinepenalty \fi}}}
\def\eqalignno#1{\panorama \tabskip=\humongous
        \halign to\displaywidth{\hfil$\displaystyle{##}$
        \tabskip=0pt&$\displaystyle{{}##}$\hfil
        \tabskip=\humongous&\llap{$##$}\tabskip=0pt
        \crcr#1\crcr}}
\begin{document}
\bibliographystyle{apsrev}

\title{Effect of Siegert's Theorem on Low-Energy Neutrino-Nucleus Interactions }

\author{A. C. Hayes and J. L. Friar}
\affiliation{Los Alamos National Laboratory, Los Alamos, NM, USA  87545}

%\date{\today}
\begin{abstract}
We examine the importance of conserving the vector current in calculating low-energy neutrino-nucleus interactions by
implicitly invoking Siegert's Theorem in describing the vector transverse electric current.
We find that at low neutrino energies ($E_\nu < $50 MeV), Siegert's Theorem can  change neutrino cross sections for normal-parity 
non-spin-flip excitations by about a factor of two. The same is true of muon capture rates.
At higher neutrino energies the effect of Siegert's Theorem diminishes, and by about 100 MeV the effect is very small. 
%%%%%%%%%%%%%%% LA-UR Rotated Box %%%%%%%%%%%%%%
%\hspace*{-1.3in}
%\rotatebox{90}{%
%\fbox{\parbox[t]{0.95in}{LA-UR-13-27154}}}
%\vspace*{-1.05in}
%%%%%%%%%%%%%%%%%%%%%%%%%%%%%%%%%%%%%%%%
\end{abstract}
\maketitle
%\pacs{28.41.-i,23.40.-s,23.40.Bw,14.60.Lm, 14.60.St}
%%%%%%%%%%%%%%%%%%%%%
\section{Introduction}
%%%%%%%%%%%%%%%%%%%%%
The use of nucleon and pion degrees of freedom in describing low-energy nuclear processes is standard in the field. This is hardly surprising since asymptotic  channels in low-energy nuclear reactions often contain nucleons, whose rest masses determine the bulk of the total energy in a nucleus. Pions on the other hand are hidden at low energies (compared to the pion mass), but their effect is extremely important. Much effort has been expended in describing the forces between nucleons in terms of two-, three-, and even four-nucleon forces generated by pion exchanges.  The effect of pion exchanges is more subtle on the various mechanisms that produce nuclear reactions. One of the best studied is meson-exchange currents (MEC), particularly in electromagnetic (EM) interactions. The exchange of charged pions can generate an EM current, which can be significant.

A long-standing discrepancy ($\approx 10\%$) between the observed and calculated thermal neutron-proton capture rates ($ n+p \rightarrow d + \gamma$) was resolved in 1970 by Riska and Brown \cite{RB}, who used explicit pion-exchange mechanisms to explain
 the discrepancy in terms of MEC. This reaction is  primarily magnetic dipole and is dominated by the large isovector magnetic moment of the nucleon, which suppresses the {\it relative} contribution of MEC. In other channels (such as electric dipole) the relative contribution can be much larger ($\approx 50\%$). Pionic MEC contributions to the nuclear Compton amplitude are needed for gauge invariance of the latter, and were calculated by Friar in 1976 \cite{FC}.

The analogous pionic MEC in the axial charge operator (also relatively large $\approx 30\%$) was calculated by Kubodera {\it et al.} in 1978 \cite{KDR} and was applied to $0^+ \rightarrow 0^-$ transitions by Haxton \cite{HA} in 1981. It is clear from the sizes of these pionic currents that either MEC should be explicitly added to calculations, or other means (viz., ``tricks'') used to incorporate MEC at some level of accuracy.

At low energies in EM interactions the latter is possible because of Siegert's Theorem (ST) \cite{ST}, a version of which can be accomplished using a vector identity. Use of ST greatly improves the interpretability of electromagnetic reactions. If one works in the long-wavelength limit for photons (real or virtual) the exponential photon wave function can be ignored, leading to a transition operator for the EM current, $\bJ (\bx)$,
$$
 \int d^3 x \, \bJ (\bx) \equiv -\int d^3 x \, \bx \, \bnabla \cdot \bJ (\bx)= i[H, \int d^3 x \, \bx \,\rho (\bx)] \, , \eqno(1)
$$
where we have used the current continuity equation $\bnabla \cdot \bJ (\bx)= -i[H, \rho (\bx)] \rightarrow -i \, \omega_{fi} \, \rho(\bx)$  to produce the last form, which is exact in this limit if the current $\bJ (\bx)$ is conserved. Note that $\omega_{fi}$ is the final nuclear energy $-$ the initial nuclear energy and $\int d^3 x \, \bx \,\rho (\bx)$ is the nuclear dipole operator, which is much easier to treat and interpret than the current. The significance of the MEC can be immediately seen by separating the strong Hamiltonian H into kinetic (T) and potential (V) parts, both roughly equal in size while opposite in sign. The former part results from the single-nucleon convection current, $\bJ_C (\bx)$, and leads immediately to $\int d^3 x \, \bJ_C (\bx) = i[T, \bD]$. The potential part  ($i[V, \bD]$) results from the MEC. Thus even if one starts with single-nucleon currents, use of the trick (replacing $\bnabla \cdot \bJ$ via the current continuity equation) forces the introduction of multi-nucleon currents, provided that there is a change of isospin (i.e., a net flow of current). Unless the latter is true, $[V, \bD] \cong 0$ for the bulk of the current.
At higher energies this vector-identity trick is only approximate because there are additional terms, and a variety of forms are possible.

Of particular interest to us is the conserved vector current (CVC) in neutrino-nucleus interactions, which allows us to follow the same path that we used in EM interactions.  
% and we simply ignore the axial current. 
In specific partial waves at low energy the MEC contributions are significant, particularly in electric dipole ($0^+ \rightarrow 1^-$) transitions. Recent results from neutrino detectors have motivated many theoretical calculations of weak processes. One of the most influential papers that provides a framework for the latter is ODW (O'Connell, Donnelly, and Walecka \cite{ODW}). Cross section formulae and rates for muon capture were derived in terms of various operator types. 

Our primary interest in this work is the vector current part of the (transverse) electric transition operator, ${{\cal{\hat{T}}}^{el}_V}$, which effects normal-parity transitions. Transverse  defines those directions orthogonal to the momentum transfer, $\bq$, while longitudinal defines directions collinear with $\bq$. It is conventional in all treatments (weak or EM) to enforce current conservation on the longitudinal current (as was done by ODW just above their Eqn.~(35)), but not necessarily for the transverse current components, which we examine here.

There are at least four different forms of the transverse electric operator in the EM literature. The first is a ``standard'' one (Eqn.~(7b) of Ref. \cite{FF}) that doesn't {\it manifest} Siegert's Theorem (i.e., have a term proportional to $\bnabla \cdot \bJ$) in the long-wavelength limit,
$$
{{\cal{\hat{T}}}^{el}_{J M}} (q) = \frac{1}{q} \int d^3 x \, \bnabla \times (j_J (qx) \bY_{JJ}^M (\hat{x})) \cdot \bJ (\bx) \, , \eqno(2)
$$
where $q$ is the magnitude of the momentum transfer (to the nucleus). The second (Eqn.~(6) of Ref.~\cite{FH})  does manifest ST, and is obtained by a simple manipulation of the vector spherical harmonics $ \bY_{J l}^M (\hat{x})$ and spherical Bessel functions $j_J (qx)$ \cite{FH} in Eqn.~(2) above. The third requires more manipulations and leads to Eqn.~(7c) of Ref.~\cite{FF}, which also manifests ST
$$
\eqalignno{
&{{\cal{\hat{T}}}^{\prime \, el}_{J M}} (q) = \frac{-i}{q\sqrt{J(J+1)}} \cr
&\int d^3 x \, Y_{J M} \left[\bnabla\cdot \bJ (\bx)\, \frac{d}{dx} (x j_J (qx)) -q^2 \, \bx  \cdot \bJ (\bx)\, j_J (qx) \right]\,  (3)}
$$
and is our preferred form after replacing $\bnabla\cdot \bJ (\bx)$ by $-i[H, \rho (\bx)]$.
There is a fourth form that has good behavior in the ST limit and near it, while exhibiting possible pathological behavior in the short-wavelength regime. The latter form was discussed by Haxton and Friar \cite{FH} and should not be used except in the long-wavelength regime.

%%%%%%%%%%%%%%%%%%%%%%%%%%%%
\section{Application to neutrino-nucleus cross sections}
%%%%%%%%%%%%%%%%%%%%%%%%%%%%
The formalism for calculating neutrino-nucleus cross sections from the output of a nuclear structure calculation 
is often taken from \cite{ODW}, which involves evaluating the matrix elements of a set of electroweak transition 
operators.
The many-body matrix elements that enter the cross section are determined from the matrix elements of
a set of single-nucleon electroweak operators 
weighted by the corresponding nuclear-structure-dependent one-body density-matrix elements (OBDMEs).
The detailed properties of this set of momentum-dependent single-nucleon electroweak operators are
 listed in the tables of Donnelly and Haxton \cite{tables}.

Our primary interest is introducing the bulk of the effect of MEC into weak cross sections at low energies by using ST without the use of {\it explicit} models of MEC. This requires modification of only part of ${{\cal{\hat{T}}}}^{el}_{JM}$ (Eqn.~(37) in the requisite Eqns.~(35)-(42) of ODW) and nothing else. Thus one can use Eqns.~(39)-(42) of ODW  for the axial currents, which are unaffected by ST,  Eqn.~(38) for ${{\cal{\hat{T}}}}^{mag}_{JM}$, and Eqns.(35)-(36) for the vector charge. In addition the weak spin-magnetization current (the second term in Eqn. (37) of \cite{ODW} and Eqn.~(22b) in \cite{tables}) is unchanged. This leaves only the nucleon convection current part of ${{\cal{\hat{T}}}}^{el}_{JM}$ (the first term in Eqn. (37) of \cite{ODW}) that we need to modify. Note that integrals are ignored in the relevant formulae of ODW and  in our equations below.

Thus we replace the 
first operator term in Eqn. (37) of \cite{ODW} 
$$
{{\cal{\hat{T}}}}^{el}_{JM} = \frac{q}{M_N}F_1^V\Delta^{\prime M}_J(\bx)\;, \eqno(4)
$$
with an operator that manifests ST, for which we use Eqn.~(3) above with $\bnabla\cdot \bJ (\bx)$ replaced by $-i\, \omega_{fi}\, \rho (\bx)$.
$$
\eqalignno{
{{\cal{\hat{T}}}^{\prime \, el}_{J M}}& (q) = \frac{F_1^V Y_{JM}}{\sqrt{J(J+1)}} \cr
&\left[ \frac{\omega_{fi}}{q} g_1(q x)  + \frac{q}{2 M_N}\left(g_2(qx) +2 j_J (qx)\, x\, \frac{\!\!\!\partial}{\partial x} \right) \right] \,  ,(5)}
$$
where
$$
g_1 (z) = z j_{J+1} (z) - (J+1) j_J (z) \, , \eqno(6)
$$
and 
$$
g_2 (z) = (J+3) j_J (z) -z j_{J+1} (z) \, . \eqno(7)
$$
These forms are  the operators describing transitions between (assumed) single-nucleon states. A simpler and more tractable form of the matrix element of the operator in parenthesis in Eqn.~(5) (viz., $g_2(qx)+ \cdots)$ can be obtained by inserting the (implicit) final ($\psi_f$) and initial ($\psi_i$) single-nucleon states around that term. Integrating-by-parts half of the derivative term (in $x$) produces
$$
\eqalignno{
&\psi_f^{\dagger} (\bx) \left(g_2(qx) +2 j_J (qx)\, x\, \frac{\!\!\!\partial}{\partial x} \right)\psi_i (\bx) \cr
&\longrightarrow x\;  j_J (qx) \left( \psi_f^{\dagger} (\bx) \frac{\;\partial \psi_i (\bx)}{\!\!\!\!\!\!\!\!\!\partial x}  -   \frac{\;\partial \psi_f^{\dagger} (\bx)}{\!\!\!\!\!\!\!\!\!\partial x}\, \psi_i (\bx)       \right)\, . \,\,\,(8)}
$$
%%%%%%%%%%%%%%%%%%%%%%%%%%%%%%%
\subsection{The cross section to the giant dipole resonance}
%%%%%%%%%%%%%%%%%%%%%%%%%%%%%%%
The transverse electric operator only produces normal-parity transitions with total changes in angular momentum 
$\Delta J^\pi =  0^+, 1^-,  2^+,  3^-, 4^+, \cdots$, while only those that do not  change spin (viz., $\Delta S=0$) but change  isospin (viz., $\Delta T =1$) involve ST. Thus, invoking ST does not lead to any change in the lowest energy contributions to neutrino cross sections that are dominated by Gamow-Teller $1^+$ transitions.
However, other multipoles are affected, particularly transitions to the giant dipole resonance (GDR).
To estimate the size of the effect on $\Delta J^\pi =1^- ,\Delta L=1, \,\Delta S=0,\, \Delta T =1$ dipole transitions, we 
consider the neutrino cross section
for a pure GDR excitation on a closed $^{16}$O core. 
We describe the GDR resonance as a single state at 22.3 MeV  of excitation 
in terms of  a harmonic oscillator $p\rightarrow sd$ cross-shell transition
 that corresponds to the SU(3) $(\lambda, \mu) =(1,0), \Delta L=1, \,\Delta S=0$ transition, and use an oscillator parameter $b=1.8$ fm.
This state contains the full E1 strength built on a closed p-shell core. 
While this represents  a very simplified description of the GDR, 
it serves as a reasonable example for the purposes of sensitivity studies of
 the importance of ST.

In Fig.~1 we show the neutral-current neutrino cross section to the GDR state defined by the $(\lambda, \mu) =(1,0), \Delta L=1, \Delta S=0$
transition. Using Eqn.~(5) rather than Eq.~(4) (i.e. implementing ST) increases the cross section at low neutrino energies by a factor of about 2.5. At higher neutrino energies
 the
effect of ST diminishes, and by about 100 MeV the effect is very small.

We also examined the ($\nu_e, e^-)$ GDR contribution to the cross section for the electron neutrinos produced in the pion 
decay-at-rest (DAR) process, for which the neutrino flux is determined by the Michel spectrum.
The predicted cross section to the GDR state increases by about 63\% when ST is included implicitly, being
 $1.58\times 10^{-42}$ cm$^2$ with ST  and
$0.97\times 10^{-42}$ cm$^2$ when ST is omitted. In addition, the shape of the cross section angular distribution
for the two cases is also different, as shown in Fig.~2.
Though the cross section to the $p\rightarrow sd$ $3^-$ S=0 state is small, it also shows sensitivity to ST, and is enhanced by about a factor of two at low neutrinos energies.  A similar factor is well known \cite{BB}  in  photonuclear reactions.
\begin{figure}
\includegraphics [width =3.5 in]{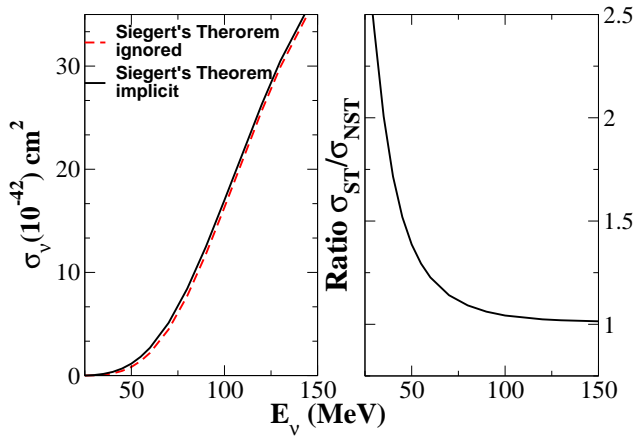}
\caption{(left) The neutral-current neutrino cross section 
for excitation of a simple GDR state at 22.3 MeV excitation energy in $^{16}$O.
The solid black curve shows the case when ST is included in the calculation and the dashed red curve when it is not.
 (right) The ratio of the ST implicit calculated cross section to the case without ST. 
The inclusion of ST increases the cross section at low neutrino energies by more than a factor of two, 
but has little effect at high neutrino energies.}
\end{figure}  
\begin{figure}
\includegraphics [width =3.5 in]{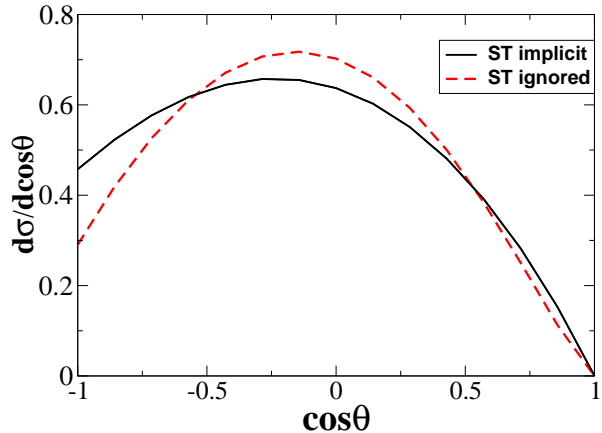}
\caption{The normalized angular distribution for the GDR contribution to the neutrino  DAR cross section in $^{16}$O.
The black curve represents a calculation where ST has been included, and the red dashed curve where it has not.
The inclusion of ST causes the angular distribution to be more forward peaked. The magnitudes of the two cross sections differ by a factor of two, being $1.2\times 10^{-42}$ cm$^2$ when ST is included implicitly, and $0.6\times 10^{-42}$ cm$^2$ when ST is omitted.}
\end{figure}
%%%%%%%%%%%%%%%%%%%%%%%%%%
\subsection{Shell model calculations for $^{12}$C}
%%%%%%%%%%%%%%%%%%%%%%%%%%
To estimate the importance of including ST in calculations of total inclusive
cross sections, we examine the neutrino cross section to the 
excited-state continuum of $^{12}$C.
For this we used the full $2\hbar\omega$ shell model calculation of ref. \cite{hayes-towner}, labeled in the latter reference as the
unrestricted shell model, which has an oscillator parameter $b=1.7$ fm.
This calculation includes correlations in both the $^{12}$C ground state and the excited states. 
The model space contain about 5500 states and includes the multipoles 
$1^+ -  5^+$ and $0^- - 4^-$, with
spurious center-of-mass states eliminated exactly.
For neutrino energies up to 100 MeV, the cross section is  dominated by the $1^+, 1^-$, and $2^-$ multipoles,
with $2^+$ and higher-order multipoles making up less than 5\% of the cross section.
Thus, to a good approximation the cross section is only affected by ST through the $\Delta S=0$ $\Delta J^\pi = 1^-$ multipole contribution, 
and to a much lesser extent, by the $\Delta S=0$ $\Delta J^\pi = 2^+$ contribution.
In Fig.~3 we show the ratio of the predicted total $(\nu_e,e^-)$ cross sections on $^{12}$C to the excited states of $^{12}$N, but
 excluding the Gamow-Teller transition to  the ground state of $^{12}$N.
As can be seen, when all multipoles are included  ST increases the total cross section by 11\% at 30 MeV and by less than 2\% at 75 MeV.
\begin{figure}
\includegraphics [width =3.5 in]{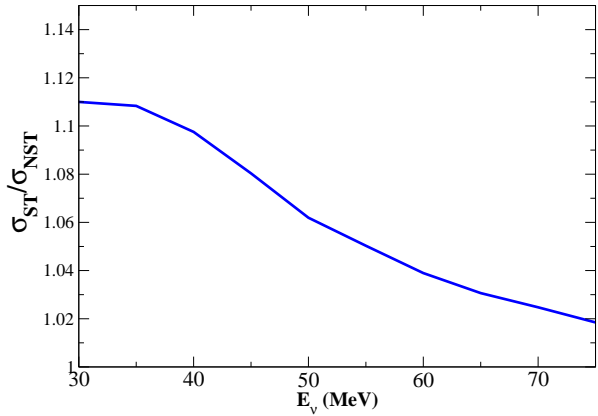}
\caption{The ratio of the total $^{12}$C($\nu_e,e^-)^{12}$N$^*$ cross section to the excited states of $^{12}$N, with and without the inclusion of ST. 
The cross sections were derived from a complete $2\hbar\omega$ shell model calculation \cite{hayes-towner} and include all multipoles $0^+ - 5^+$ and $0^- - 4^-$. They do not include the transition to the $^{12}$N$_{g.s.}$.
}
\end{figure}

%%%%%%%%%%%%%%%%%%%%%%%%%%%
\section{Muon capture to exclusive states in $^{12}$B}
%%%%%%%%%%%%%%%%%%%%%%%%%%%
The effect of ST on positive-parity contributions to neutrino cross sections is difficult to observe
because at low neutrino energies, where ST has the most impact, $2^+$ or higher multipoles contribute little to the
 total cross section.
However, muon capture rates to individual excited states of the final nucleus are measurable.
Recently, the Double Chooz (DC) collaboration measured \cite{Doublechooz} the products of $\mu$-capture on several light nuclei, using the DC neutrino detector designed to measure the neutrino mixing angle $\theta_{13}$.

We use the full $2\hbar\omega$ shell model calculation discussed above to examine the effect of ST on the 
$^{12}$C$(\mu^-, \nu_\mu)^{12}$B capture rates.   
The results are listed in Table 1.
As can be seen, invoking ST can change the muon capture rates to low-lying 2$^+$ states by more than a factor of two.
We note that these calculations use harmonic oscillator wave functions. In reference \cite{hayes-towner} the use of more
realistic radial wave functions was found to lower the predicted muon capture rates to $^{12}$B. However, the sensitivity of
the predictions with more realistic radial wave functions to ST would likely be similar to the results shown here.

\begin{table}
\caption{Muon capture rates to the low-lying states of $^{12}$B from the $^{12}$C$(\mu^-,\nu_\mu)$ reaction, in units
of 10$^{-3}$ sec$^{-1}$. 
The column labeled ST implicitly includes Siegert's theorem (Eqn.~(5)) and that labeled NST ignores Siegert's theorem (Eqn.~(4)). 
The experimental values are taken from \cite{Doublechooz}.
}
\begin{tabular}{llll}
&&&\\\hline\hline
\;\;State\;\;& ST & NST& Experiment\\
&&&\\\hline
1$^+$ (g.s.) &\;\;5.3\;\;&\;\;5.3\;\;&\;\; 5.68$^{+0.14}$$\hspace{-.7cm}_{-.23}$\;\;\\
2$^+$$_1$ (0.953 MeV)&\;\;0.167\;\;&\;\;0.441\;\;&\;\; 0.321$^{+0.09}$$\hspace{-.7cm}_{-.07}$\;\;\\
2$^-$$_1$ (1.674 MeV)&\;\;0.136\;\;&\;\;0.136\;\;&\;\; 0.06$^{+0.04}$$\hspace{-.7cm}_{-.03}$\;\;\\
1$^-$$_1$ (2.621 MeV)&\;\;0.98\;\;&\;\;1.75\;\;&\;\; 0.47$^{+0.06}$$\hspace{-.7cm}_{-.05}$\;\;\\
2$^+$$_2$ (3.759 MeV)&\;\;0.021\;\;&\;\;0.029\;\;&\;\; 0.026$^{+0.015}$$\hspace{-.7cm}_{-.011}$\;\;\\
& & & \\\hline
\end{tabular}
\end{table}

%%%%%%%%%%%%%%%%%
\section{Conclusion} 
We have shown that if CVC is not included in the calculation of ${{\cal{\hat{T}}}}^{el}_{JM}$ for neutrino-nucleus interactions at low energies, errors on the order of 
a factor of two are possible for multipoles that are normal parity and non-spin flip. 
In such cases, CVC can be invoked using Siegert's Theorem, which results in the vector transverse electric transition operator of the form given by Eqn. (3). 
Neutrino processes that are affected included those
involving neutrino energies typical of Michel pion decay-at-rest spectra, supernova neutrinos, as well as muon capture.

%%%%%%%%%%%%%%%%%

\end{document}
\la f | \int d^3 x \, \bJ (\bx) | i \ra = - \la f | \int d^3 x \, \bx \, \bnabla \cdot \bJ (\bx) | i \ra = i \omega \la f | \int d^3 x \, \bx \rho (\bx) | i \ra
{{\cal{\hat{T}}}^{\prime \, el}_{J M}} (q) = \frac{-i}{q\sqrt{J(J+1)}} \int d^3 x \, Y_{J M} [\bnabla\cdot \bJ (\bx) g(qx) -q^2 \bx  \cdot \bJ (\bx) j_J (qx) ]\, , \eqno(1)
\bibitem{daya-bay} F. P. An {\it et al.}, {\it Phys. Rev. Lett.} {\bf 108}, 171803 (2012).{\color{magenta} CHECKED}
\bibitem{reno}  J. K. Ahn {\it et al.}, {\it Phys. Rev. Lett.} {\bf 108}, 191802 (2012).{\color{magenta} CHECKED}
\bibitem{Doublechooz} Y. Abe {\it et al.}, {\it Phys. Rev. Lett.} {\bf 108}, 131801 (2012).{\color{magenta} CHECKED}
\end{document}